# Comment: Fisher Lecture: Dimension Reduction in Regression

**Ronald Christensen**

I am pleased to participate in this well-deserved recognition of Dennis Cook's remarkable career.

Cook points out Fisher's insistence that predictor variables in regression be chosen without reference to the dependent variable. Reduction by principal components clearly satisfies that dictum. One of my primary objections to partial least squares regression when I first encountered it as an alternative to principal components was that the predictor variables were being chosen with reference to the dependent variable. (I now have other objections to partial least squares.) Yet on the other hand, variable selection in regression is well accepted and it clearly chooses variables based on their relationship to the dependent variable. Perhaps variable selection is better thought of as a form of shrinkage estimation rather than as a process for choosing predictor variables.

Cook also reiterates something that I think is difficult to overemphasize: Fisher's point that "More or less elaborate forms [for models] will be suitable according to the volume of the data." We see this now on a regular basis as modern technology provides larger data sets to which elaborate models are regularly fitted.

With regard to Cook's work, it seems to me that the key issue in the development of Cook's models (2), (5), (10) and (13) is whether they are broadly reasonable. The question did not seem to be extensively addressed but Cook shows that much can be gained if we can reasonably use them. When they are appropriate, the results in the corresponding propositions are rather stunning. It has long been known that the best regression model available—technically the best predictor of a random variable $y$ based on a $p$-dimensional random vector $x$—is the conditional mean $\mathrm{E}(y|x)$. The problem with this result is that it requires us to know the joint distribution of $(x', y)$. Most of what we commonly recognize as regression analysis is an attempt to model the relationship $\mathrm{E}(y|x)$. This includes linear regression, nonlinear regression, generalized linear models and the various approaches to "nonparametric" (actually, highly parametric) regression. Under the models being considered, there exists a $p \times d$ matrix $\Gamma$ such that

$$y|x \sim y|\Gamma'x.$$

This means that $\mathrm{E}(y|x) = \mathrm{E}(y|\Gamma'x)$ regardless of what modeling strategy we choose to use. If anything, this dimensionality reduction from $p$ to $d$ is of more importance to nonparametric regression than other forms because, as the number of predictor variables increases, nonparametric regression gets hit harder by the curse of dimensionality than less highly parametric forms. As a result, nonparametric regression should benefit most from the existence of a generally valid reduction in dimensionality.

The issue with these four models is to estimate the column space of $\Gamma$, say, $C(\Gamma)$. In the first six sections, the results are all closely tied to the eigenvectors (principal component vectors) of some estimated covariance matrix for the predictor variables $x$, say $\hat{\Sigma}$. For model (2), the space is spanned by the first $d$ principal component vectors of the usual $\hat{\Sigma}$. For model (5), the space is spanned by the first $d$ principal component vectors of a restricted version of $\hat{\Sigma}$. For models (10) and (13), the estimation procedure is a bit more complicated. The key is that for both models (10) and (13) the population covariance matrix of $x$ can be written as

$$\Sigma = \Gamma V D V' \Gamma + \Gamma_0 V_0 D_0 V_0' \Gamma_0,$$

with $D$ and $D_0$ diagonal matrices, in such a way that

$$\Sigma(\Gamma V) = (\Gamma V)D, \quad \Sigma(\Gamma_0 V_0) = (\Gamma_0 V_0)D_0.$$

*Ronald Christensen is Professor, Department of Mathematics and Statistics, University of New Mexico, Albuquerque, New Mexico 87131-0001, USA e-mail: fletcher@stat.unm.edu.*

This implies that the eigenvectors of $\Sigma$ are either in $C(\Gamma)$ or in $C(\Gamma_0) \equiv C(\Gamma)^\perp$, the orthogonal complement of $C(\Gamma)$. The problem is to establish which $d$ out of the $p$ orthogonal eigenvectors belong in $C(\Gamma)$. To estimate $C(\Gamma)$, find the orthogonal eigenvectors of $\hat{\Sigma}$, say, $v_1, \ldots, v_p$, and check the likelihood of every one of the $p$ choose $d$ combinations that has $d$ of the $v_i$'s in $C(\Gamma)$ and the remaining $p-d$ vectors in the orthogonal complement. Whichever combination maximizes the likelihood provides the estimate of $C(\Gamma)$. In case $\binom{p}{d}$ is large, Cook provides a sequential selection method. The key difference between the procedures for models (10) and (13) is that the likelihoods are different.

The remainder of my discussion is an attempt to put the question of estimating the reduced space into the context of multivariate linear model theory. To do this, I will change Cook's notation completely, so that the problem looks more like standard multivariate linear models, but then reidentify the parts of the problem that interest Cook. I do not presume that any of this is new to Cook, but I found it helpful in understanding the process.

In discussing multivariate linear models, liberal use is made of Kronecker products, Vec operators, and their properties; see, for example, Christensen (2002, Definition B.5 and Subsection B.5). Recall also that for a univariate linear model $Y = X\beta + e$, $\mathrm{E}(e) = 0$ and $\mathrm{Cov}(e) = V$,

$$SSE \equiv (Y - X\hat{\beta})'V^{-1}(Y - X\hat{\beta})$$
$$= Y'V^{-1}(Y - X\hat{\beta}).$$

Moreover, least squares estimates are BLUE's, and thus maximum likelihood, if $C(VX) \subset C(X)$. We will apply these facts to the multivariate models. Finally, let $J_r^s$ denote an $r \times s$ matrix of 1's with $J_r \equiv J_r^1$, and for a matrix $A$ let $P_A = A(A'A)^- A'$ be the perpendicular projection operator (ppo) onto $C(A)$ with $\mathrm{r}(A)$ the rank of $A$.

The standard multivariate linear model involves dependent variables $y_1, \ldots, y_q$. If $n$ observations are taken on each dependent variable, we have $y_{ih}$, $i = 1, \ldots, n$, $h = 1, \ldots, q$. Let $Y_h = [y_{1h}, \ldots, y_{nh}]'$ and $y_i' = [y_{i1}, \ldots, y_{iq}]$. For each $h$, we have a linear model,

$$Y_h = X\beta_h + e_h,$$
$$\mathrm{E}(e_h) = 0,$$
$$\mathrm{Cov}(e_h) = \sigma_{hh} I,$$

where $X$ is a known $n \times p$ matrix that is the same for all dependent variables, but $\beta_h$ and the error vector $e_h = [e_{1h}, \ldots, e_{nh}]'$ are peculiar to the dependent variable $Y_h$.

The multivariate linear model consists of fitting the $q$ linear models simultaneously. Letting

$$Y_{n \times q} = [Y_1, \ldots, Y_q],$$
$$B_{p \times q} = [\beta_1, \ldots, \beta_q],$$
$$e_{n \times q} = [e_1, \ldots, e_q],$$

the multivariate linear model is

$$(1) \qquad Y = XB + e.$$

Alternatively, thinking of $X$ as a matrix with rows $x_i'$ and $e$ as having rows $\varepsilon_i'$, we can write the multivariate linear model as

$$y_i' = x_i' B + \varepsilon_i', \quad i = 1, \ldots, n.$$

To perform maximum likelihood, we assume that the $\varepsilon_i$'s are independent $N(0, \Sigma)$ random vectors. It is reasonably well known that for any ppo $P_A$,

$$(2) \qquad \mathrm{E}_{Y|X}(Y' P_A Y) = \mathrm{r}(A)\Sigma + B'X'P_A XB.$$

$\Sigma$ is now being used for the conditional covariance matrix of $y|x$, whereas Cook used $\Sigma$ for the marginal covariance matrix of $x$.

A generalization of the multivariate linear model that is often associated with growth curve models (cf. Christensen, 2001) is

$$(3) \qquad Y = X\Gamma Z' + e,$$

where the unknown parameter matrix $B$ in (1) is replaced by a product of a reduced parameter matrix $\Gamma$ that is $p \times d$ and a fixed, known matrix $Z$ that is $q \times d$ with $\mathrm{r}(Z) \leq d < q$. This is essentially Cook's model (5) when applied to data and using drastically different notation. (Our $Y$ is his $X$, our $X$ is his known function of $y$, $F_y$, our $Z$ is his $\Gamma$, etc.) The ultimate goal of our exercise is to drop the assumption that we know $Z$ and estimate it, or more properly $C(Z)$, from the data. But for now, we act as if $Z$ is known. Note that the "growth curve" model specifies something akin to a linear model for each row of $Y$,

$$y_i = Z(\Gamma x_i) + \varepsilon_i, \qquad i = 1, \ldots, n.$$

Moreover, using Kronecker products and Vec operators, we can turn the multivariate growth curve model (3) into a univariate linear model,

$$(4) \qquad \mathrm{Vec}(Y) = [Z \otimes X]\mathrm{Vec}(\Gamma) + \mathrm{Vec}(e),$$

with
$$\mathrm{Vec}(e) \sim N(0, [\Sigma \otimes I_n]).$$

There are a couple of refinements to model (4) used in Cook's development. First, the growth curve model is specified as

(5) $$Y = J\mu' + X\Gamma Z' + e,$$

with $J'X = 0$ and $Z'Z = I_d$. As a linear model (5) becomes
$$\mathrm{Vec}(Y) = [I_q \otimes J_n]\mu + [Z \otimes X]\mathrm{Vec}(\Gamma) + \mathrm{Vec}(e).$$

Second is the assumption in Cook's models (2) and (5) that
$$\Sigma = \sigma^2 I_q,$$

in which case
$$\mathrm{Cov}[\mathrm{Vec}(e)] = \sigma^2[I_q \otimes I_n] = \sigma^2 I_{nq},$$

so standard estimation results apply to the model. In particular, least squares estimates of the parameters $\mu$ and $\mathrm{Vec}(\Gamma)$ are maximum likelihood estimates and the likelihood function for fixed $\sigma^2$ evaluated at the maximum likelihood estimates of $\mu$ and $\Gamma$ is, ignoring the constant,

(6) $$-\frac{nq}{2}\log(\sigma^2) - \frac{SSE}{2\sigma^2}.$$

Performing the usual computations necessary to finding least squares estimates, but using properties of Kronecker products and Vec operators and exploiting the fact that since $J'X = 0$ we have $C([I_q \otimes J_n]) \perp C([Z \otimes X])$ so that estimation of $\mu$ and $\Gamma$ can be performed separately, the least squares estimates reduce to
$$\hat{\mu} = \bar{y}_{\cdot\cdot}, \quad \hat{\Gamma} = (X'X)^- X'YZ(Z'Z)^-,$$
or, alternatively,
$$X\hat{\Gamma}Z = P_X Y P_Z.$$

The maximum likelihood estimate of $\sigma^2$ is obtained by differentiating (6) with respect to $\sigma^2$ and setting it equal to 0, yielding $\hat{\sigma}^2 = SSE/nq$. In particular, the perpendicular projection operator onto $C([I_q \otimes J_n], [Z \otimes X])$ is $[I_q \otimes (1/n)J_n^n] + [P_Z \otimes P_X]$, so

$$SSE = \mathrm{Vec}(Y)'[I_q \otimes (I - (1/n)J_n^n)]\mathrm{Vec}(Y)$$
$$\quad - \mathrm{Vec}(Y)'[P_Z \otimes P_X]\mathrm{Vec}(Y)$$
$$= \|\mathrm{Vec}[(I - (1/n)J_n^n)Y]\|^2$$
(7)
$$\quad - \|\mathrm{Vec}(P_X Y P_Z)\|^2$$
$$= \mathrm{tr}[Y'(I - (1/n)J_n^n)Y]$$
$$\quad - \mathrm{tr}[P_Z Y' P_X Y P_Z].$$

Using notation analogous to Cook's, three estimators that we will use frequently are
$$\hat{\Sigma} \equiv \frac{1}{n}Y'\left(I - \frac{1}{n}J_n^n\right)Y,$$
$$\hat{\Sigma}_{\mathrm{fit}} \equiv \frac{1}{n}Y'P_X Y,$$
$$\hat{\Sigma}_{\mathrm{res}} \equiv \frac{1}{n}Y'\left(I - \frac{1}{n}J_n^n - P_X\right)Y.$$

Note that $\hat{\Sigma}$ is the maximum likelihood estimate of the covariance matrix when fitting the usual multivariate one-sample model $Y = J_n\mu' + e$. Using the assumption $Z'Z = I_d$ so that $ZZ' = P_Z$,
$$SSE = \mathrm{tr}[n\hat{\Sigma}] - \mathrm{tr}[Z'n\hat{\Sigma}_{\mathrm{fit}}Z].$$

As Cook points out, this depends on $C(Z)$ rather than $Z$ itself.

We are finally in a position to address Cook's question, the fact that we do not actually know $Z$. To maximize the likelihood (6) as a function of $Z$ we need to maximize $\mathrm{tr}[Z'n\hat{\Sigma}_{\mathrm{fit}}Z]$ as a function of $Z$ subject to $Z'Z = I_d$. If we think about finding the columns of $Z$ sequentially, that is, finding $z_1$ to maximize $z'\hat{\Sigma}_{\mathrm{fit}}z$ subject to $\|z_1\|^2 = 1$, then finding $z_2$ to maximize $z'\hat{\Sigma}_{\mathrm{fit}}z$ subject to $\|z_2\|^2 = 1$ and $z_2'z_1 = 0$, and so on, this is a standard problem in multivariate analysis that is solved by finding the eigenvectors of $\hat{\Sigma}_{\mathrm{fit}}$ relative to the eigenvalues $\lambda_1 \geq \lambda_2 \geq \cdots \geq \lambda_q \geq 0$. Of course, since $Z$ is $q \times d$, we consider only the first $d$ eigenvectors.

To examine a model equivalent to Cook's model (2), we consider the most extreme (largest) choice for $X$, which is $C(X) = C(J_n)^\perp$. In this case, $P_X = I_n - (1/n)J_n^n$ so that $\hat{\Sigma}_{\mathrm{fit}} = \hat{\Sigma}$. It follows that the maximum likelihood estimate of $Z$ consists of the first $d$ principal component vectors. As pointed out by Cook, the number of parameters in our $Z$ matrix is $pd$. However, with this choice of $X$, $p = n - 1$, so the number of parameters rises with the sample size.

For Cook's models (10) and (13) in Section 6, the covariance structure changes. As indicated earlier, the estimation methods ultimately involve determining which of the principal component directions are most likely where principal components can be computed from some estimate of $\Sigma$, which may be any, or preferably all, of $\hat{\Sigma}$, $\hat{\Sigma}_{\mathrm{fit}}$ or $\hat{\Sigma}_{\mathrm{res}}$. For Cook's model (13) we again have
$$\mathrm{Vec}(Y) = [I_q \otimes J_n]\mu + [Z \otimes X]\mathrm{Vec}(\Gamma) + \mathrm{Vec}(e),$$

but recalling that $Z'Z = I_d$, we now incorporate a matrix $Z_0$ with $Z_0'Z = 0$ and $Z_0'Z_0 = I_{q-d}$ and assume

$$\text{Vec}(e) \sim N(0, [Z_0\Omega_0^2 Z_0' + Z\Omega^2 Z' \otimes I_n]).$$

Observe that least squares estimates will still be BLUEs and thus maximum likelihood estimates because $C([Z_0\Omega_0^2 Z_0' + Z\Omega^2 Z' \otimes I_n][Z \otimes X]) \subset C([Z \otimes X])$.

The *SSE* now involves the perpendicular projection operators as in (7), but also involves the inverse of the covariance matrix. With our assumptions about $Z$ and $Z_0$,

$$[Z_0\Omega_0^2 Z_0' + Z\Omega^2 Z' \otimes I_n]^{-1}$$
$$= [Z_0\Omega_0^{-2} Z_0' + Z\Omega^{-2} Z' \otimes I_n].$$

The *SSE* becomes

$$SSE = \text{Vec}(Y)'[(Z\Omega^{-2} Z' + Z_0\Omega_0^{-2} Z_0')$$
$$\otimes (I - (1/n)J_n^n)] \text{Vec}(Y)$$
$$- \text{Vec}(Y)'[Z\Omega^{-2} Z' \otimes P_X] \text{Vec}(Y)$$
$$= \text{Vec}(Y)' \text{Vec}[(I - (1/n)J_n^n)$$
$$\cdot Y(Z\Omega^{-2} Z' + Z_0\Omega_0^{-2} Z_0')]$$
$$- \text{Vec}(Y)' \text{Vec}(P_X Y Z\Omega^{-2} Z')$$
$$= \text{tr}[Y'(I - (1/n)J_n^n)Y(Z\Omega^{-2} Z' + Z_0\Omega_0^{-2} Z_0')]$$
$$- \text{tr}(Y' P_X Y Z\Omega^{-2} Z')$$
$$= \text{tr}[\Omega_0^{-2} Z_0' Y'(I - (1/n)J_n^n)Y Z_0]$$
$$+ \text{tr}\{\Omega^{-2} Z'[Y'(I - (1/n)J_n^n)Y - Y' P_X Y]Z\}$$
$$= \text{tr}[\Omega_0^{-2} Z_0' n\hat{\Sigma} Z_0] + \text{tr}\{\Omega^{-2} Z'[n\hat{\Sigma} - n\hat{\Sigma}_{\text{fit}}]Z\}.$$

The likelihood will be

$$\frac{-n}{2} \log(|Z_0\Omega_0^2 Z_0' + Z\Omega^2 Z'|) + \frac{-1}{2} SSE$$
$$= \frac{-n}{2} \log(|\Omega_0^2|) + \frac{-n}{2} \text{tr}[\Omega_0^{-2} Z_0' \hat{\Sigma} Z_0]$$
$$+ \frac{-n}{2} \log(|\Omega^2|) + \frac{-n}{2} \text{tr}(\Omega^{-2} Z'[\hat{\Sigma} - \hat{\Sigma}_{\text{fit}}]Z),$$

which, maximizing over $\Omega$ and $\Omega_0$, Cook indicates reduces to

$$\frac{-n}{2} \log(|Z_0' \hat{\Sigma} Z_0|) + \frac{-n}{2} \log(|Z'[\hat{\Sigma} - \hat{\Sigma}_{\text{fit}}]Z|).$$

As before, Cook's model (10) can be viewed as the special case where $C(X) = C(J_n)^\perp$, so that the second term in the likelihood disappears.

To continue this discussion, we need to leave the conditional world of linear models and consider the unconditional expected values of $\hat{\Sigma}$, $\hat{\Sigma}_{\text{fit}}$ and $\hat{\Sigma}_{\text{res}}$. Conditionally, applying (2) to model (5) when $J'A = 0$ gives

(8) $\quad \text{E}_{Y|X}(Y' P_A Y) = \text{r}(A)\Sigma + Z\Gamma' X' P_A X \Gamma Z'.$

For $\hat{\Sigma}$ and $\hat{\Sigma}_{\text{fit}}$ the appropriate ppo has $P_A X = X$ and for $\hat{\Sigma}_{\text{res}}$ the ppo has $P_A X = 0$. We have assumed that $J'X = 0$, which is only reasonable if the random rows of $X$ have been adjusted by their sample means; nonetheless, it is reasonable to define the marginal covariance matrix of a row of $X$ as

$$V_x \equiv \frac{1}{n-1} \text{E}_X(X'X).$$

These results quickly yield the expectations

$$\text{E}(\hat{\Sigma}) = \frac{n-1}{n}\Sigma + \frac{n-1}{n} Z\Gamma' V_x \Gamma Z',$$

$$\text{E}(\hat{\Sigma}_{\text{fit}}) = \frac{\text{r}(X)}{n}\Sigma + \frac{n-1}{n} Z\Gamma' V_x \Gamma Z',$$

$$\text{E}(\hat{\Sigma}_{\text{res}}) = \frac{n-1-\text{r}(X)}{n}\Sigma.$$

In particular, with $\Sigma = Z_0\Omega_0^2 Z_0' + Z\Omega^2 Z'$, Cook's Proposition 4 says that the estimates converge in probability to the limits of their expected values.

Cook's second simulation has a true model with $d = 1$, $n = 250$, $q = 10$ (his $p$), $p = 1$, $\Gamma = 1$, $\Omega = \sigma$, $\Omega_0 = \sigma_0 I_9$ and $V_x = \sigma_x^2$ (his $\sigma_Y^2$). Here,

$$\text{E}(\hat{\Sigma}) = \frac{n-1}{n}\sigma_0^2 Z_0 Z_0' + \frac{n-1}{n}(\sigma^2 + \sigma_x^2) Z Z',$$

$$\text{E}(\hat{\Sigma}_{\text{fit}}) = \frac{\text{r}(X)}{n}\sigma_0^2 Z_0 Z_0'$$
$$+ \left(\frac{\text{r}(X)}{n}\sigma^2 + \frac{n-1}{n}\sigma_x^2\right) Z Z',$$

$$\text{E}(\hat{\Sigma}_{\text{res}}) = \frac{n-1-\text{r}(X)}{n}\sigma_0^2 Z_0 Z_0'$$
$$+ \frac{n-1-\text{r}(X)}{n}\sigma^2 Z Z'.$$

Cook's simulation results make good sense in terms of these expected values. Terms involving $\text{r}(X)/n$ should be unimportant. When $\sigma_0$ is small, $\text{E}(\hat{\Sigma})$ is dominated by $(\sigma^2 + \sigma_x^2)ZZ'$, which is larger than the corresponding terms $\sigma_x^2 ZZ'$ and $\sigma^2 ZZ'$ for $\hat{\Sigma}_{\text{fit}}$ and $\hat{\Sigma}_{\text{res}}$, respectively, so it works best. When $\sigma_0$ is comparable to $\sigma$ and $\sigma_x$, $\hat{\Sigma}_{\text{fit}}$ works well, because $\text{E}(\hat{\Sigma}_{\text{fit}})$ is much less affected by $\sigma_0^2$ than the other estimates. And when $\sigma_0$ is large, $\hat{\Sigma}_{\text{res}}$ works well because then we need to look at the eigenvectors for small eigenvalues of $\hat{\Sigma}_{\text{res}}$ and $\hat{\Sigma}$, but the term $\sigma^2 ZZ'$ for $\hat{\Sigma}_{\text{res}}$ is smaller than the term $(\sigma^2 + \sigma_x^2)ZZ'$ for $\hat{\Sigma}$, whereas the expected value of $\hat{\Sigma}_{\text{fit}}$ is relatively unaffected by $\sigma_0^2$ getting large. As Cook mentions, when $\sigma_x^2 + \sigma^2 = \sigma_0^2$, there is very little ability for $\hat{\Sigma}$ to identify $C(Z)$ because then $\text{E}(\hat{\Sigma}) = (n-1)\sigma_0^2/nI_q$, so we cannot really expect the eigenvectors of $\hat{\Sigma}$ to help us identify $C(Z)$. Similarly, when $\sigma^2 = \sigma_0^2$, $\text{E}(\hat{\Sigma}_{\text{res}}) = (n - 1 - \text{r}(X))\sigma_0^2/nI_q$.

Cook's models (2), (5), (10) and (13) involve specialized structure for $\Sigma$. His model (16) allows for general $\Sigma$. Nonetheless, the expectations of the estimates show that there should almost always be some ability to estimate $C(Z)$. In particular,

$$\text{E}\left[\frac{n}{\text{r}(X)}\hat{\Sigma}_{\text{fit}} - \frac{n}{n-1}\hat{\Sigma}\right] = \frac{n-1-\text{r}(X)}{\text{r}(X)} Z\Gamma'V_x\Gamma Z',$$

so the first $d$ principal component vectors of the estimate $\frac{n}{\text{r}(X)}\hat{\Sigma}_{\text{fit}} - \frac{n}{n-1}\hat{\Sigma}$ should be at least a reasonable estimate of a basis for $C(Z)$. For large samples this is similar to looking at the directions determined by $\hat{\Sigma}_{\text{fit}}$, but in the extreme case of $C(X) = C(J_n)^\perp$, the estimator is degenerate at 0. Of course, according to Cook's Proposition 6, for the general $\Sigma$ (Cook's $\sigma^2\Delta$) of model (16), it no longer suffices to estimate $C(Z)$; we need to estimate $C(\Sigma^{-1}Z)$. Fortunately, $\hat{\Sigma}_{\text{res}}$ provides an estimate of $\Sigma$, so we can just transform the estimated basis for $C(Z)$. For large $n$, it makes sense to base estimation of $C(Z)$ on $\hat{\Sigma}_{\text{fit}}$, but rather than transforming its eigenvectors, one could alternatively look directly at the eigenvectors of $\hat{\Sigma}_{\text{res}}^{-1}\hat{\Sigma}_{\text{fit}}$, which is Cook's recommendation when $p = d$. This intuitive approach based on the expected values of (matrix) quadratic forms seems analogous to using Henderson's method 3 for estimating variance components, whereas Cook is recommending a maximum likelihood procedure, which I suspect is better.

I found the relationship between model (16) and ordinary least squares (OLS), discussed in Section 7.4, fascinating. It seems that the gains to be had over OLS demonstrated in the simulations are due to imposing alternative structure on the inverse regression relationship. I am comforted by the idea that using OLS is not bad but rather, if we can find an appropriate model with more structure, we can do better than OLS. Of course, this only applies when the sufficient reduction is one-dimensional. With more dimensions, we need our full range of regression tools to develop a relationship between the dependent variable and the reduced predictor variables.

I initially found Cook's discussion of standardization in Section 7.3 disturbing. I am not dogmatic about the need to standardize variables prior to finding principal components. When the measurements are all on similar scales, using the original scales seems reasonable to me, as when measuring the height, length and width of turtle shells in centimeters. On the other hand, if I measure length in kilometers and height and width in millimeters, the first principal component will essentially ignore the lengths, regardless of any role that length might play in prediction. I suspect that one point of Cook's discussion is that in a situation where you need to standardize the variables, there will be little reason to suppose that his models (2) or (5) are appropriate, which means there is little reason to use principal components. More generally, his point seems to be that standardization is necessary but that a more complete standardization than merely rescaling the variables is needed.

As I indicated at the beginning of my discussion, my biggest problem with these procedures is that I do not have a good feel for when the models (2), (5), (10) and (13) will be appropriate. Multivariate linear model theory should allow us to use $\hat{\Sigma}_{\text{res}}$ to test the assumption of Cook's models (2) and (5) that $\Sigma = \sigma^2 I$. I am less sure if it will provide a test of whether $\Sigma = \sigma^2 ZZ' + \sigma_0^2 Z_0 Z_0'$, when $Z$ is unknown, but a generalized likelihood ratio test seems plausible. In any case, the procedures in Section 7.2 seem generally applicable.